\documentclass[conference]{IEEEtran}
\usepackage{cite}
\usepackage{enumerate}
\usepackage{graphics}
\usepackage{caption3}
\usepackage{commath}
\usepackage[dvipsnames]{xcolor}
\usepackage{comment}
\usepackage{amsmath,amssymb,amsfonts}
\usepackage{algorithmic}
\usepackage{graphicx}
\usepackage{mathtools}
\usepackage{siunitx}
\usepackage{booktabs,array}
\usepackage{epsfig}
\usepackage{epstopdf}
\usepackage{listings}
\usepackage[caption=false]{subfig}
\usepackage{soul}
\usepackage{textcomp}
\def\BibTeX{{\rm B\kern-.05em{\sc i\kern-.025em b}\kern-.08em
    T\kern-.1667em\lower.7ex\hbox{E}\kern-.125emX}}

\usepackage[acronym,shortcuts,nonumberlist,nopostdot,nogroupskip]{glossaries}
\makeglossaries
\newacronym{5g}{5G}{fifth-generation}
\newacronym{adoa}{ADoA}{angle difference-of-arrival}
\newacronym{aoa}{AoA}{angle of arrival}
\newacronym{aod}{AoD}{angle of departure}
\newacronym{ap}{AP}{access point}
\newacronym{brp}{BRP}{beam refinement protocol}
\newacronym{cdf}{CDF}{cumulative distribution function}
\newacronym{cir}{CIR}{channel impulse response}
\newacronym{cots}{COTS}{commercial off-the-shelf}
\newacronym{csi}{CSI}{channel state information}
\newacronym{dnn}{DNN}{deep neural network}
\newacronym{los}{LoS}{line of sight}
\newacronym{mpc}{MPC}{multipath component}
\newacronym{mimo}{MIMO}{multiple-input multiple-output}
\newacronym{mmw}{mmWave}{millimeter wave}
\newacronym{mse}{MSE}{mean-square error}
\newacronym{mmse}{MMSE}{minimum mean-square error}
\newacronym{nlos}{NLoS}{non-line of sight}
\newacronym{nn}{NN}{neural network}
\newacronym{noma}{NOMA}{non-orthogonal multiple access}
\newacronym{rss}{RSS}{received signal strength}
\newacronym{rssi}{RSSI}{received signal strength indicator}
\newacronym{slam}{SLAM}{simultaneous localization and mapping}
\newacronym{sls}{SLS}{sector-level sweep}
\newacronym{snr}{SNR}{signal-to-noise ratio}
\newacronym{ssw}{SSW}{sector sweep}
\newacronym{tdoa}{TDoA}{time difference-of-arrival}
\newacronym{toa}{ToA}{time of arrival}
\newacronym{tof}{ToF}{time of flight}
\newacronym{va}{VA}{virtual anchor}
\newacronym{ar}{AR}{augmented reality}
\newacronym{vr}{VR}{virtual reality}
\newacronym{wlan}{WLAN}{wireless LAN}

\begin{document}

\bstctlcite{references:BSTcontrol}

\title{Millimeter Wave Localization with Imperfect Training Data using Shallow Neural Networks%
}

\author{\IEEEauthorblockN{Anish Shastri}
\IEEEauthorblockA{\textit{DISI, University of Trento}\\
Trento, Italy \\
anish.shastri@unitn.it\vspace{-1mm}}
\and
\IEEEauthorblockN{Joan Palacios}
\IEEEauthorblockA{\textit{North Carolina State University}\\
Raleigh, NC, USA \\
jbeltra@ncsu.edu\vspace{-1mm}}
\and
\IEEEauthorblockN{Paolo Casari}
\IEEEauthorblockA{\textit{DISI, University of Trento}\\
Trento, Italy \\
paolo.casari@unitn.it\vspace{-1mm}}
}

\maketitle

\begin{abstract}

\Ac{mmw} localization algorithms exploit the quasi-optical propagation of \ac{mmw} signals, which yields sparse angular spectra at the receiver. 
Geometric approaches to angle-based localization typically require to know the map of the environment and the location of the access points. Thus, several works have resorted to automated learning in order to infer a device's location from the properties of received \ac{mmw} signals. However, collecting training data for  such models is a significant burden.
In this work, we propose a shallow neural network model to localize \ac{mmw} devices indoors. This model requires significantly fewer weights than those proposed in the literature. Therefore, it is amenable for implementation in resource-constrained hardware, and needs fewer training samples to converge.
We also propose to relieve training data collection efforts by retrieving (inherently imperfect) location estimates from geometry-based \ac{mmw} localization algorithms.
Even in this case, our results show that the proposed neural networks perform as good as or better than state-of-the-art algorithms.
\end{abstract}

\begin{IEEEkeywords}
Millimeter wave; AoA; ADoA; Neural networks; Indoor localization;
\end{IEEEkeywords}

\glsresetall

\section{Introduction}

\Ac{mmw} communication technology in the 30--300~GHz band is showing great promise for high-data rate, short-range wireless communications~\cite{mmwaveNYU}. A number of applications are starting to benefit from the capabilities of \ac{mmw} technology: these applications include \ac{ar}, \ac{vr}~\cite{mmwaveApps}, indoor robot navigation, as well as asset tracking in Industry~4.0 scenarios.

\acp{mmw} attenuate strongly over much shorter distances than sub-6~GHz signals. Hence, \ac{mmw} equipment resorts to tightly packed antenna arrays and directive beam patterns in order to improve the received power levels. The arrays focus emitted signals towards pre-defined directions, and limit the number of reflections of the signal on the surrounding environment. Such effect compounds with the quasi-optical propagation patterns of \ac{mmw} signals, which reflect crisply off indoor surfaces and obstacles with limited scattering, just like light rays~\cite{mmWaveNYU5g}. 
As a result, received \ac{mmw} signals have a significantly sparse angular spectrum, typically composed of one \ac{los} and multiple \ac{nlos} \acp{mpc}. These features enable effective angle-based localization approaches.

Angle-based localization schemes relieve some shortcomings of other approaches based, e.g., on \ac{rssi} or \ac{snr} information, which require to find an optimal, environment-specific mapping between such metrics and the communication distance.
In particular, geometry-based schemes such as~\cite{palacios2017jade} and~\cite{clam2018Palacios}  
map \ac{nlos} \acp{mpc} to \acp{va}, thereby exploiting multipath propagation to find the location of a device.
Such location systems achieve good accuracy and do not require any initial knowledge about the location of the anchors and reflective surfaces in an indoor environment. 
However, they typically need several iterations to compute a single location estimate, as well as to refine estimates as the device collects fresh measurement data. Yet, such refinement steps become more complex as more data is input to the algorithm. 
Thus, such schemes may not be amenable to implementation in real-time on embedded devices.

To counter some of the above issues, other contributions employ \acp{dnn} in order to localize a device~\cite{deepL2020mitsubishi, wangMERLfingerprintingPart4}. These works are usually \ac{ap}-centric, meaning that the \ac{ap} collects measurements and localizes the clients. Such algorithms usually require the knowledge of the environment, and do not scale easily. They typically have two drawbacks: a)~they need a large dataset to train the \ac{dnn} models, akin to fingerprinting techniques that require the radio map of the entire indoor environment; b)~large \ac{dnn} models provide accurate location estimates, but can be computationally very complex.



In this paper, we tackle the above issues by proposing the use of shallow \acp{nn} to solve the user device localization problem in an indoor environment. These \ac{nn} models have far fewer weights than \acp{dnn}, and adapt much better to energy-constrained or computationally limited platforms. 
Moreover, instead of explicitly collecting well-labeled and calibrated training data, we propose to collect a training dataset by running an angle-based localization algorithm such as JADE~\cite{palacios2017jade}. 
This relieves the effort-intensive training data collection phase, by associating angle information to locations computed by JADE, at the cost of producing error-prone location labels. In spite of this, we show that our shallow \ac{nn} model can still achieve a satisfactory degree of accuracy, comparable to that of models trained with perfect data.

Unlike \ac{dnn} algorithms for \ac{mmw} localization from the literature, our proposed algorithm is device-centric, that is, localization is carried out by each device in a distributed fashion. This makes it more scalable for two reasons: i) each device autonomously estimates its own location; and ii) different devices can cooperate to the collection of training data for the shallow models, e.g., by uploading them to a training server, and receive a trained \ac{nn} model, after which each device becomes fully independent. 
We show that this approach leads to accurate localization by achieving sub-meter accuracy in up to 90\% of the cases, in rooms of different shape and with a typical number of \acp{ap} deployed.  

\noindent The specific contributions of our work are:
\begin{itemize}
    \item A shallow neural network model that can estimate the coordinates of the user device in an indoor environment by leveraging \ac{aoa} measurements for \acp{mpc} in receiver-side angular spectra; 
    \item The training of our shallow \acp{nn} with imperfect location estimates from an angle-based localization algorithm, before switching to them for location estimation: this avoids the explicit collection of training data;
    \item A simulation campaign that assesses the performance of our algorithm against a state-of-the-art scheme from the literature, in two different indoor deployments.
\end{itemize}
For the latter, we also discuss the impact of the training dataset size on the accuracy of our \acp{nn}, showing that our shallow models yield satisfactory accuracy even with a limited number of training samples.

The remainder of this paper is organized as follows: Section~\ref{sec:related} presents a summary of the literature on indoor \ac{mmw} localization; Section~\ref{sec:propalgo} describes our proposed algorithm; Section~\ref{sec:simresults} presents simulation results in two different indoor environments; finally, we draw our conclusions in Section~\ref{sec:conclusion}.

\section{Related work}
\label{sec:related}

In this section, we survey indoor localization approaches based on \ac{mmw} technology. We subdivide these works into classical and machine learning-based localization schemes.

\subsection{Classical mmWave localization systems}

Most of the schemes proposed in the literature exploit \ac{mmw} signal attributes such \ac{aoa}, \ac{csi}, and \ac{rssi} to localize the user device in an indoor environment. For example, in~\cite{olivier2016lightweight,palacios2019single}, the authors present triangulation and \ac{adoa}-based schemes for device localization. These schemes require the knowledge of a device's orientation, of the surrounding environment, and of the \ac{ap} deployment information. In~\cite{accurate3D2018pefkianakis}, the \ac{cir} of the received \ac{mmw} signals makes it possible to estimate \ac{aoa} and \ac{tof} information, and thereby localize a device in 3D. However, the approach in~\cite{accurate3D2018pefkianakis} also requires the knowledge of the indoor environment. 

\ac{ap}-centric localization algorithms such as~\cite{leap2019palacios} exploit \ac{csi} measurements to infer angle information from \ac{mmw} signals sent by a client. 
A map-assisted positioning technique is proposed in~\cite{kanhere2019map} to estimate the location of the user. The authors simulate the technique on data collected at 28 and 73~GHz through a 3D ray tracer. 
The device localization and environment mapping technique in~\cite{clam2018Palacios} uses the beam training procedure to acquire \ac{aoa} information, employs \acp{adoa} to localize both a \ac{mmw} device and all physical and virtual anchors in the environment, and simultaneously maps the environment's boundaries. The scheme is experimentally evaluated on 60 GHz \ac{mmw} hardware, albeit not on \ac{cots} devices.

\subsection{Machine learning-based mmWave localization systems}

Several works employed machine learning and deep learning to localize \ac{mmw} devices indoors. For example, Vashist \emph{et al}.{} resort to a multi-layer perceptron regression model in order to estimate the coordinates of a client~\cite{Vashist2020ml}. The model uses SNR values as fingerprints to localize the agent in 1D.
Pajovic \emph{et al}.{} started from \ac{rssi} and beam indices fingerprint datasets and designed probabilistic models to estimate the client location~\cite{fing2019mitsubishi1}. An extension of the same work~\cite{fing2019mitsubishi2}, uses spatial beam \ac{snr} datasets for position and orientation classification, as well as coordinate estimation. Deep learning techniques are also the main enablers for localization in~\cite{deepL2020mitsubishi} and~\cite{wangMERLfingerprintingPart4}, where the authors proposed ResNet-inspired models~\cite{resnet2015deep} for device localization in \ac{los} and \ac{nlos} scenarios.
To tackle the challenges imposed by \ac{nlos} conditions, the authors use spatial beam \ac{snr} values in~\cite{deepL2020mitsubishi}, whereas they employ multi-channel beam covariance matrix images in~\cite{wangMERLfingerprintingPart4}.

\subsection{Summary}

The above discussion shows that experimentally-validated \ac{mmw} localization schemes tend to have one of the following shortcomings: i) low-complexity geometric techniques (e.g., based on triangulation~\cite{palacios2019single}) require to know the orientation of each device and the map of the environment; moreover they are not robust against erroneous \ac{aoa} estimates; ii) geometric schemes that collect multiple \ac{aoa} measurements and progressively refine location estimates experience increasingly higher complexity when the number of collected measurements becomes significant; iii) deep learning-based techniques rely on fingerprinting, and imply a significant training data collection burden (especially in large and challenging environments), can be lengthy to train, and may require excessive computational resources.

In contrast to the above, our proposed \ac{nn} models are shallow, thus less complex to run and faster to train. To collect ground truth data for \ac{nn} training, we do not carry out preliminary measurements. Rather, we resort to a localization algorithm from the literature, and switch to the \ac{nn} model after accruing a sufficient number of location estimates to train the model successfully. While such location labels are inherently affected by estimation errors, we still show that our models achieve a good level of accuracy even in these conditions.


\section{Proposed localization scheme}
\label{sec:propalgo}

In this section, we explain the proposed algorithm in detail. We first explain the idea behind the proposed technique and then explain the network architecture and its components.

\subsection{Main Idea}
We propose to use a shallow regression \ac{nn} model (with two hidden layers and one output layer) in order to learn the relationship between \ac{adoa} measurements collected by a \ac{mmw} client and the locations at which these measurements were collected.
Being based on regression rather than classification, the proposed \ac{nn} model is more robust to error-prone training data. This makes it possible to train the \ac{nn} using imperfect location labels, as would be obtained, e.g., from a geometry-based \ac{mmw} localization algorithm.
Our approach thus greatly reduces both the \ac{nn} training data collection effort and the complexity of accurate geometry-based \ac{mmw} localization algorithms, that tends to increase with the number of collected \ac{adoa} measurements.

\subsection{Input features}
\begin{figure}[t]
\centering
\includegraphics[width=0.5\columnwidth]{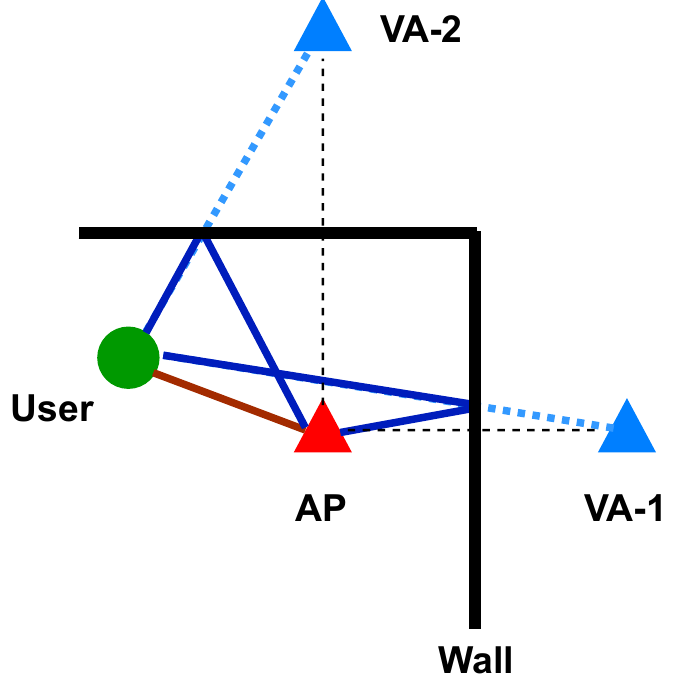}
 \caption{Illustration of the concept of virtual anchor and first-order reflections.}
 \label{fig:VAdiag}
\end{figure}

We rely on angle information to estimate the location of a \ac{mmw} client. In particular, we exploit the sparseness of the angular power spectra of the signal that the client receives from visible \acp{ap} in order to distinguish different \acp{mpc} and measure the \ac{aoa} for each of them. Due to multipath propagation, \acp{mpc} can be either \ac{los} or \ac{nlos} paths that have undergone one or more reflections. Since second- and higher-order reflections bear comparatively lower power than \ac{los} and first-order \ac{nlos} \acp{mpc}, we neglect \acp{mpc} with more than one reflection in our approach.
We remark that \acp{aoa} from \ac{nlos} paths can be directly mapped to the \acp{va}, which appear as mirror images of physical \acp{ap} with respect to each reflective surface in the indoor area (e.g., walls)~\cite{palacios2017jade}. Fig.~\ref{fig:VAdiag} illustrates how \acp{va} can be modeled as the (virtual) source of \ac{nlos} paths, and how \ac{los} and \ac{nlos} \acp{mpc} correspond to different \acp{aoa} due to multipath propagation. In the following we collectively refer to \acp{ap} along with their corresponding \acp{va} as \emph{anchors}. 

After measuring the angular spectrum from each \ac{ap} at the client, we elect one reference \ac{mpc} and compute \acp{adoa} with respect to the \ac{aoa} of such \ac{mpc}. If we collect \ac{aoa} measurements from $N_a$ different anchors, we therefore obtain $N_a - 1$ \ac{adoa} values. Using \acp{adoa} values makes the location estimation problem invariant to the client orientation.
We employ these \acp{adoa} as the input features of our \ac{nn} model.

\subsection{Neural network architecture}

\begin{figure}[t]
\centering
\includegraphics[width=0.8\columnwidth]{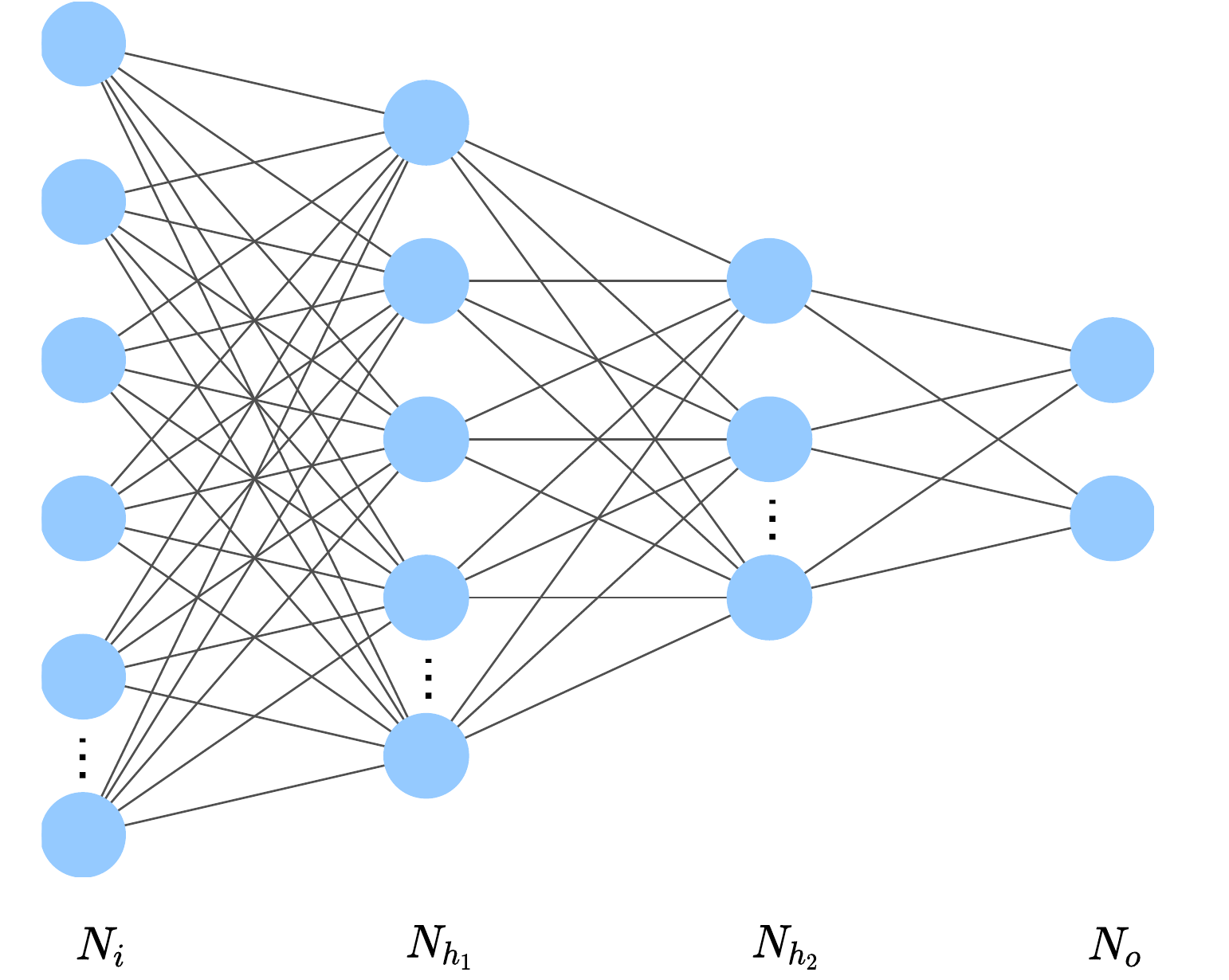}
 \caption{Structure of our proposed NN model.}
 \label{fig:nnModel}
\end{figure}

Our proposed neural networks consist of three layers: the input layer has $N_i = N_a-1$ neurons; the first hidden layer contains $N_{h_1}$ neurons; the second hidden layer $N_{h_1}/2$ neurons; finally, the output layer consists of 2 neurons, one for each of the 2D coordinates of the client. 
Fig.~\ref{fig:nnModel} shows the general structure of the neural network, whereas Table~\ref{nnModelparam} summarizes the structure of the number of neurons in each layer. Here, $\lceil{\cdot}\rceil$ denotes the ceiling function.

\begin{table}[t]
\caption{Summary of the number of neurons in each layer}
\renewcommand{\arraystretch}{1.1}
\centering
\begin{tabular}{@{\hspace{1mm}}lc@{\hspace{1mm}}}
\toprule
 \textbf{Layer} & \textbf{Number of neurons} \\

\midrule
 Input layer~($N_i$) & $N_a - 1$  \\
 Hidden layer 1~($N_{h_1}$) & $\lceil{N_i \cdot k \rceil}$ \\
 Hidden layer 2~($N_{h_2}$)  & $\lceil N_{h_1}/2 \rceil $\\
 Output layer~($N_o$) & $2$ \\
\bottomrule
\end{tabular}
\label{nnModelparam}
\end{table}

We train our model to learn the non-linear function that maps the input \ac{adoa} data to the output client coordinates. 
Call ${\bf x}$ the location of the client, and denote the input \ac{adoa} vector at location ${\bf x}$ as $\mathbf{a}_{\bf x}$, and the weight matrix of layer $i$, ${\bf W}_i$, as
\begin{equation}
    {\bf W}_i = [{\bf w}_1 \cdots {\bf w}_{n_i}] \; ,
\end{equation}
where $n_i$ is the number of neurons in layer $i$, and ${\bf w}_j$ is the $n_{i-1} \times 1$ column vector containing the weights of each link that connects the $n_{i-1}$ neurons in the previous layer to the $j$th neuron in the current layer. Moreover, let ${\bf b}_i = [b_1 \cdots b_{n_i}]^T$ be the vector of bias values for each of the neurons in layer $i$, and call the neuron activation function $A(\cdot)$. Therefore, the output values of layer $i$ are
\begin{equation}
    \mathbf{y}_i = A({\bf W}_i^T {\bf y}_{i-1} + \mathbf{b}_i) \; ,
\end{equation}
where with a small abuse of notation, applying $A(\cdot)$ to a vector denotes applying $A(\cdot)$ to each entry of the vector. Note that ${\bf y}_0 = {\bf a}_{\bf x}$.
Because our \acp{nn} have $n=3$ layers, the estimated location of the client is
\begin{equation}
    \hat{\bf x} = [\hat{x}_1,\hat{x}_2]^T = \mathbf{y}_3 = \mathcal{F}({\bf a}_{\bf x}) \; , 
\end{equation}
where
\begin{equation}
    \mathcal{F}({\bf a}_{\bf x}) = A\Big( {\bf W}_3^T A\big( {\bf W}_2^T A({\bf W}_1^T{\bf a}_{\bf x}+{\bf b}_1) + {\bf b}_2\big) +{\bf b}_3 \Big)
\end{equation}
is the non-linear regression function applied by the \ac{nn}. Given the very small number of neurons in our \acp{nn}, computing ${\cal F}(\cdot)$ only requires a few simple matrix multiplications and vector summations, making the operation affordable even for computationally-constrained devices. 
In this work, we choose the rectified linear activation function (ReLU) for $A(\cdot)$, and train the network via the Adam optimizer. 
As we cast the regression problem as a \ac{mse} minimization problem, we employ the \ac{mse} loss function
\begin{equation}
    L({\bf x},{\hat{\bf x}}) = |x_1 - \hat{x}_1|^2 + |x_2 - \hat{x}_2|^2
\label{regressionLoss}   
\end{equation}
where $(x_1,x_2)$ are the true location of the user and $(\hat{x}_1,\hat{x}_2)$ are the location estimates output by the \ac{nn}.

We remark that our \ac{nn}s implement a regression model, i.e., they estimate the coordinates of the client location, rather than identifying which position among multiple possible choices best corresponds to the input data. 

\subsection{Hyperparameter tuning}

Through hyperparameter tuning we can optimize the neural network model in order to minimize the loss function. In our model, we tune the following hyperparameters: i) the node factor $k$ which sets the number of neurons in the hidden layers; ii) the dropout rate $p$, which helps avoid overfitting; and iii) the learning rate $r$, which controls the initial speed of convergence. Table~\ref{hyperparameters} summarizes the hyperparameters used for network model optimization. For the learning rate, we consider a logarithmic progression of values from $10^{-4}$ up to $10^{-2}$, with ten values per decade.

\begin{table}[t]
\caption{Summary of the hyperparameters chosen to tune our model}
\renewcommand{\arraystretch}{1.1}
\centering
\begin{tabular}{@{\hspace{1mm}}lc@{\hspace{1mm}}}
\toprule
 \textbf{Hyperparameter} & \textbf{Range} \\
\midrule
 Node factor~($k$) & $\{0.6,~0.7,~0.8\}$  \\
 Dropout rate ($p$) & $\{0\%,~5\%,~10\%\}$ \\
 Learning rate ($r$) & $[0.0001, 0.01]$\\
\bottomrule
\end{tabular}

\label{hyperparameters}
\end{table}

\section{Simulation results}
\label{sec:simresults}


\subsection{Simulation environment and setup}

\begin{figure}[t]
    \centering
    \subfloat[Rectangular room.\label{fig:rectRoom}]{\includegraphics[width=0.49\columnwidth]{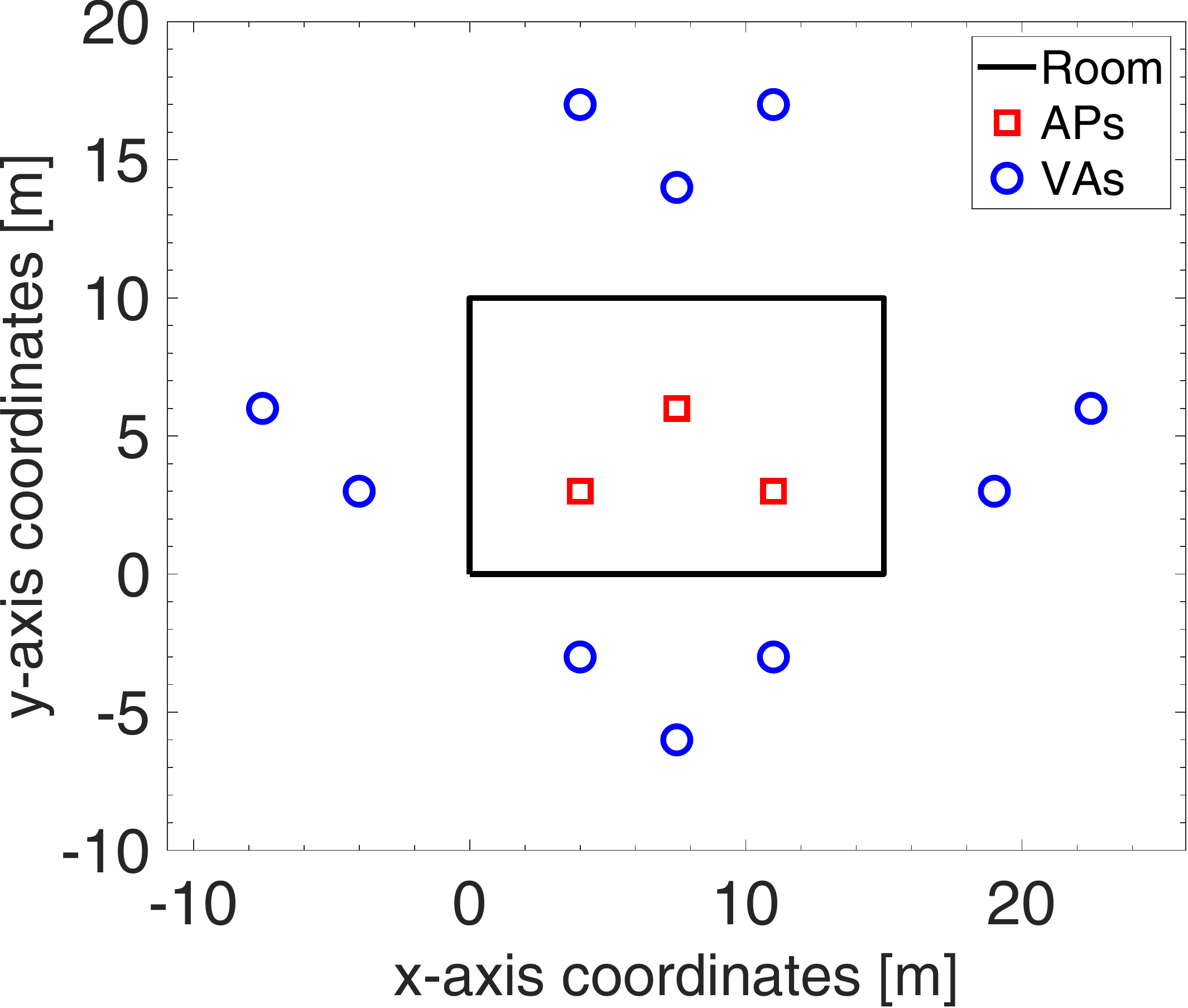}}
    \hfill
    \subfloat[Reverse L-shaped room\label{fig:Lroom}]{\includegraphics[width=0.49\columnwidth]{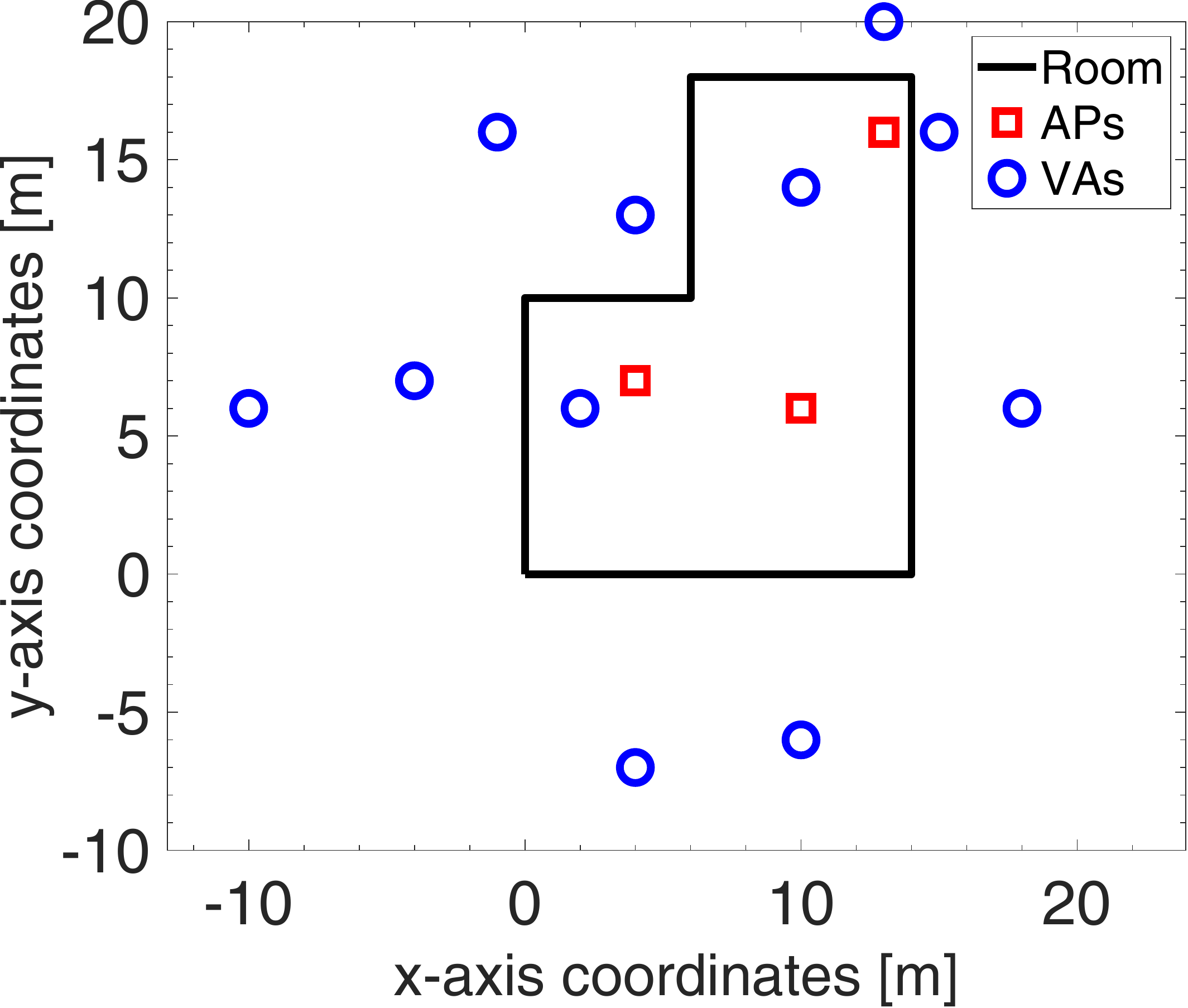}}
    \caption{Simulation scenarios showing the shape of two rooms, and the location of APs and VAs.}
    \label{fig:scenarios}
\end{figure}

We test our scheme in two indoor environments. The first is a 15$\times$10~m$^\textrm{2}$ rectangular room as shown in Fig.~\ref{fig:rectRoom}, where three \ac{mmw} access points (red squares) are deployed at the coordinates $(4,3)$, $(7.5,6)$ and $(11, 3)$, where all values are in meters. 
Each \ac{ap} also maps to 4 \acp{va} (blue circles), each corresponding to a different wall.

The second indoor environment is a reverse L-shaped room, whose bottom-left section has size 6$\times$10~m$^\textrm{2}$, whereas the right tall section has size 8$\times$18~m$^\textrm{2}$. Fig.~\ref{fig:Lroom} illustrates such L-shaped room. Three \acp{ap}, placed at coordinates $(4,7)$, $(10,6)$, and $(13,16)$ cover the entire indoor space. As before we depict most \acp{va} as blue circles. 

To produce simulation results that are as close to realistic scenarios as possible, we consider the \ac{aoa} measurements to be affected by zero-mean Gaussian noise of standard deviation error in the set $\{ 5^{\circ}, 7^{\circ}, 10^{\circ} \}$. These values are realistic in mobile user scenarios, given the imperfect beam training procedures and non-pencil-shaped beam patterns of existing commercial off-the-shelf hardware~\cite{steinmetzer2017compressive}.

To generate a training set for the \acp{nn}, we simulate several mobile client trajectories in each room, and use a ray tracer to collect \ac{aoa} information at 30~points along each trajectory. Each \ac{aoa} is affected by a Gaussian-distributed error as explained above. To compare our approach to other algorithms from the literature, we generate a completely different test set of 30~trajectories, for a total of 900~client locations. Note that the train and test dataset are uniformly generated across the entire room.



\subsection{Performance evaluation}

\begin{figure}[t]
     \centering
     \subfloat[Rectangular room with 3 APs\label{fig:boxPlotAlgos3}]{\includegraphics[width=1\columnwidth,trim={0 70mm 0 0},clip]{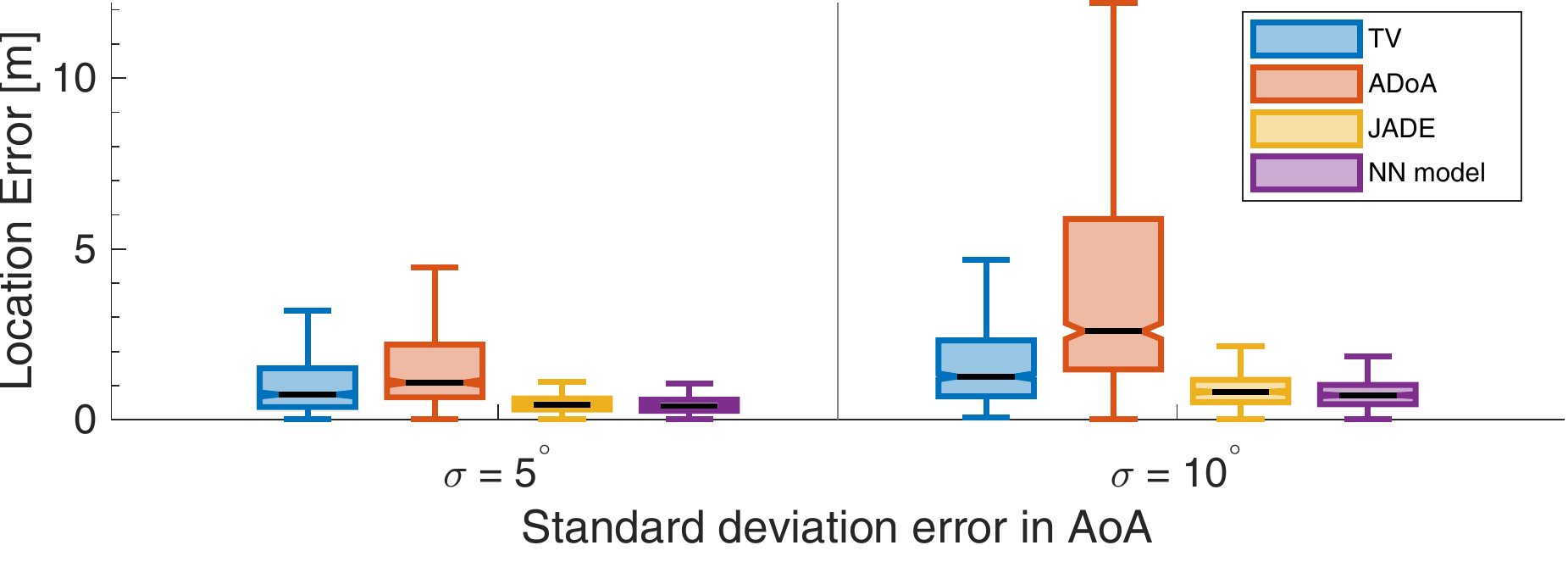}}\\
     \vspace{1mm}
     \subfloat[Rectangular room with 4 APs\label{fig:boxPlotAlgos4}]{\includegraphics[width=1\columnwidth,trim={0 0 0 70mm},clip]{allAlgoBoxplotALLAP_debug.pdf}}
     \caption{Statistical dispersion of the localization error for different values of $\sigma$ in the rectangular room with (a) 3 APs and (b) 4 APs.}
     \label{fig:boxPlotAlgos}
 \end{figure}

We start our evaluation by comparing the localization performance of our proposed approach against the JADE~\cite{palacios2017jade}, Triangulate-Validate (TV), and plain geometric \ac{adoa} algorithms~\cite{palacios2019single} from the literature, as they are state-of-the-art approaches relying only on \ac{adoa} measurements for location estimation. Fig.~\ref{fig:boxPlotAlgos} shows this comparison for the rectangular room with 3~\acp{ap} (top panel) 
as well as with 4~\acp{ap}, where in the latter case the coordinates of the \acp{ap} are $(2,1)$, $(4,7)$, $(10,4)$ and $(14,9)$.
The figure conveys the statistical dispersion of the location estimation errors for two different values of the standard deviation of \ac{aoa} errors, $\sigma=5^{\circ}$ and $\sigma=10^{\circ}$. Each box extends from the 1st to the 3rd quartile, the notch denotes the median of the distribution, and the whiskers cover the range from the 10th to the 90th percentiles.

We observe that the dispersion is small both for our \ac{nn} model and for JADE, whereas TV and \ac{adoa} yield larger errors, which increase for higher values of $\sigma$. This is a direct consequence of the geometric operations of the two algorithms, which rely on accurate \ac{aoa} measurements in order to estimate locations. 
Instead both JADE and our \ac{nn} model compensate well for the erroneous angle measurements. However, we remark that the accuracy of JADE comes from having processed a large number of measurements. As this requires to solve minimum mean-square error problems, processing exceedingly many measurements would be very time-consuming. In fact, its complexity increases both with the number of measurements and with the number of \acp{ap}~\cite{palacios2017jade}. Instead, our model learns the mapping between \ac{adoa} data and the location of the client, and its complexity remains constant after the training phase (it requires very few multiplications and additions), while achieving equivalent or better accuracy than JADE.
With 4~\acp{ap} (panel b), the general observations remain the same, except that all algorithms yield better estimation errors, with smaller maximum errors, and less statistical dispersion.
Because JADE and our \ac{nn} model outperform TV and \ac{adoa} from~\cite{palacios2019single}, in the following we will exclude the latter two algorithms from the evaluation.

\begin{figure}[t]
    \centering
    \includegraphics[width=0.95\columnwidth]{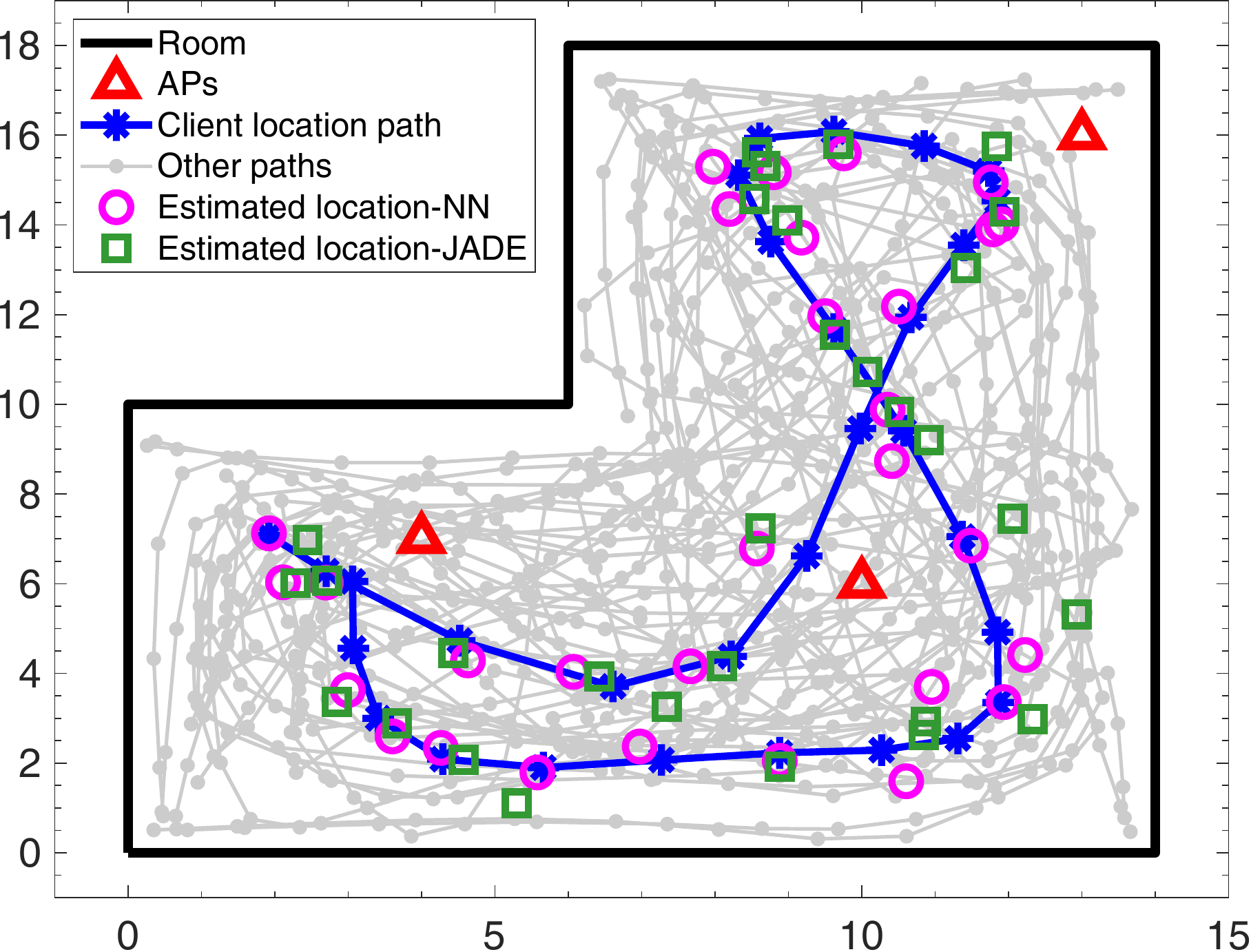}
    \caption{Path reconstruction in the L-shaped room using our neural network model and JADE. $\sigma = 5^{\circ}$. Axis tick labels are in meters.}
    \label{fig:userTrajectory}
\end{figure}

Fig.~\ref{fig:userTrajectory} shows the reconstruction of a mobile user trajectory using JADE and our \ac{nn} model, this time in the reverse L-shaped room. The global set of paths considered for our evaluation is shown through grey lines, whereas a thicker blue line conveys the path under consideration. Purple circles convey the estimates of our \ac{nn} model, whereas green squares represent JADE's location estimates. For this evaluation, \acp{aoa} are affected by errors of standard deviation $\sigma=5^{\circ}$. The resulting \ac{nn} model has $(18,13,7,2)$ neurons in each layer, and the optimal hyperparameters are $p = 0.05$ and $r = 0.002$, yielding an \ac{mse} of 0.186.
We observe that both techniques estimate the locations of the client and reconstruct the corresponding trajectory with good accuracy. The few outliers that remain, both for NN and JADE, are due to erroneous \ac{aoa} values. Still, our \ac{nn} model computes estimates closer to the ground-truth trajectory than JADE.

\begin{figure}[t]
\centering
\includegraphics[width=1\columnwidth]{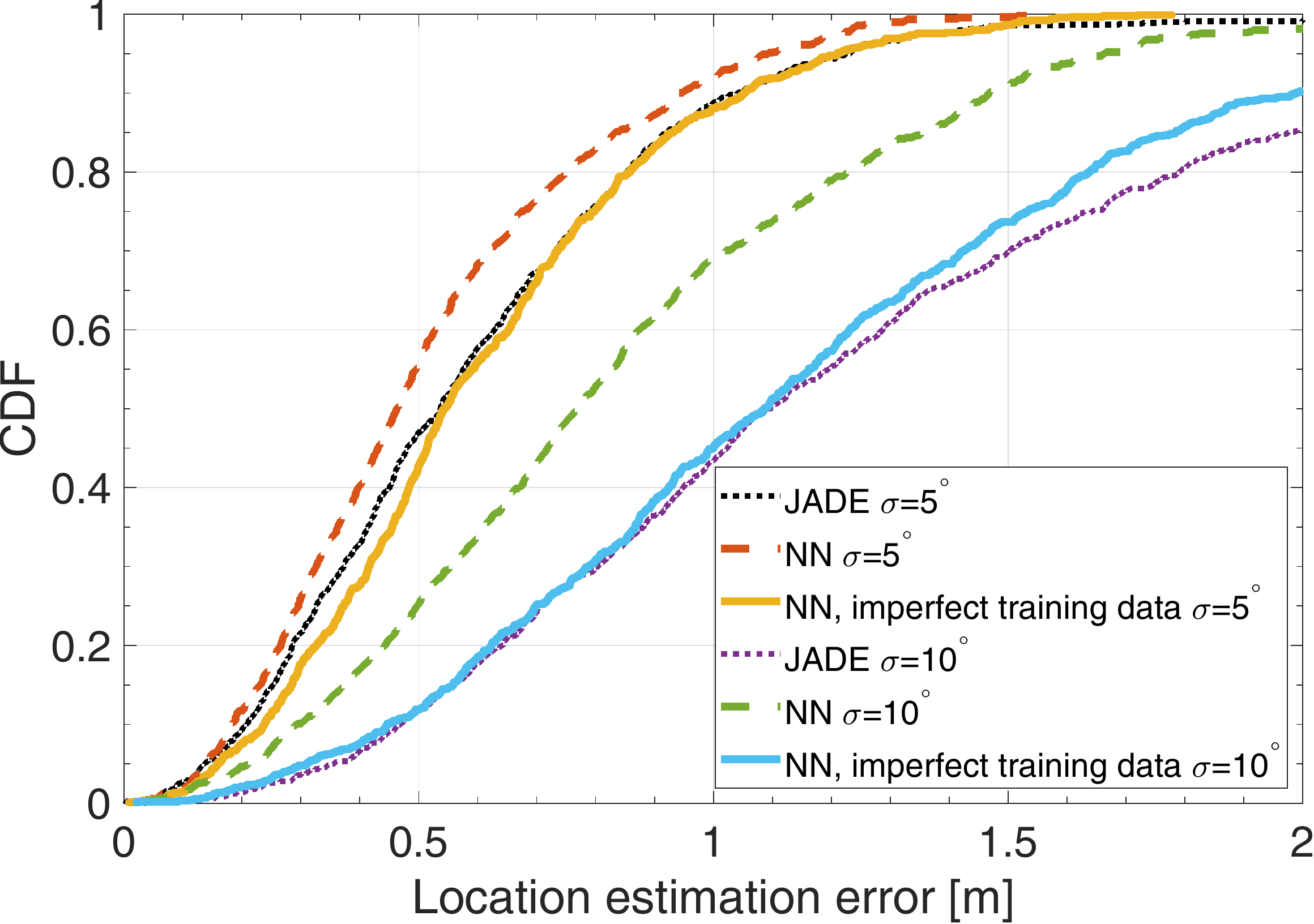}
 \caption{Cumulative distribution function of the location error for our \ac{nn} approach against JADE in the L-shaped room environment. The \acp{nn} are trained both with perfect and with error-prone data.}
 \label{fig:cdfLroomEstimatedUser}
\end{figure}

\subsection{Training the NNs with estimated user locations}

We now consider the case where we train our \ac{nn} with imperfect location labels from the JADE algorithm, which is inherently error-prone. 
The performance of the proposed method is illustrated in Fig.~\ref{fig:cdfLroomEstimatedUser} through the \ac{cdf} of the location error for different approaches, evaluated in the L-shaped room.
The \acp{nn} trained with imperfect labels have $(18,13,7,2)$ neurons in each stage. For $\sigma=5^\circ$, the optimal hyperparameters are $p = 0$ and $r = 0.006$, whereas $p = 0.05$ and $r = 0.0004$ for $\sigma = 10^{\circ}$.

We observe that the \acp{nn} trained with JADE estimates (solid blue and yellow curves) perform just as good as JADE when $\sigma=5^{\circ}$, and even better if $\sigma=10^{\circ}$. This supports our intuition that collecting error-prone training samples does not significantly deteriorate the location estimates of our \ac{nn} models. In particular, for $\sigma=5^\circ$, training the \acp{nn} with imperfect data from JADE yields sub-meter errors in 88\% of the cases, and a median error of 0.47~m. This compares very well to the 92\% sub-meter errors and the median error of 0.45~m achieved with perfectly labeled training data.

\subsection{Impact of number of training samples}

\begin{figure}[t]
\centering
\includegraphics[width=0.95\columnwidth]{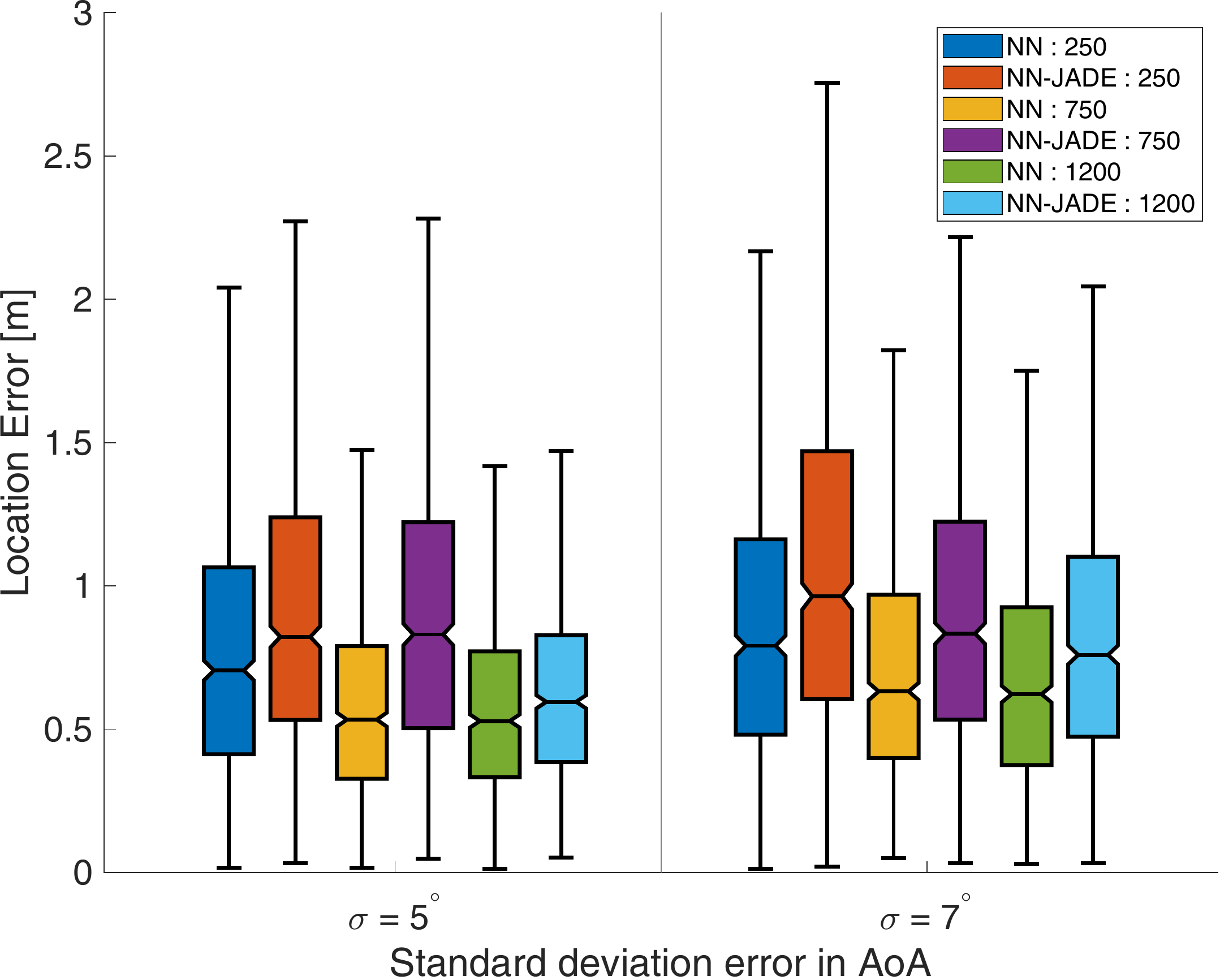}
 \caption{Statistical dispersion of the localization error for different number of training samples and values of $\sigma$, when the \ac{nn} is trained with true user locations (key: NN) and with estimates from JADE (key: NN-JADE).}
 \label{fig:boxPlotDiffsamples}
\end{figure}

We conclude our study by assessing the performance of our model as a function of the training set size. As presented in~\cite{palacios2017jade}, the complexity of JADE increases as the \ac{mmw} client keeps collecting angular spectrum measurements. Thus, we can run JADE up to a point where its complexity does not become excessive. To validate this with our proposed scheme, we consider training datasets including 250, 750, and 1200 samples. We then train our model both with perfect locations labels and with imperfect location estimates from JADE.
Fig.~\ref{fig:boxPlotDiffsamples} compares the corresponding box plots for $\sigma=5^{\circ}$ and $\sigma=7^{\circ}$. The results show that both the median location error and the statistical dispersion of the error decrease by increasing the number of training samples. This is true even when the model is trained with JADE's estimates, as such estimates become more accurate when computed from a larger set of measurements~\cite{palacios2017jade}. 
This confirms that our \ac{nn} model can be trained with a significantly small number of training samples compared to \ac{dnn} models, and that the resulting errors are acceptable even when training samples are error-prone. 

We remark that two trends emerge from the figure. On the one hand, with a comparatively small training set (blue and red boxes), the \ac{nn} may underfit the map between angular spectra and the device location: in this regime, training with imperfect estimates only marginally increases the localization error of the \ac{nn}.
On the other hand, for a large training dataset (green and light-blue boxes), the environment is well sampled, and JADE's estimates are quite accurate. Therefore, training with error-prone location labels yields almost the same accuracy as training with true locations.
Between these two regimes, training with true locations (yellow box) expectedly yields better performance than training with JADE's location estimates (purple box).
The system designer may therefore trade off localization performance with the number of samples retrieved from JADE before switching to the \ac{nn} model.



\section{Conclusions and future work}
\label{sec:conclusion}

In this paper, we proposed a shallow neural network model that localizes a client in an indoor environment. The model uses \ac{adoa} values as input and can be trained through error-prone client location estimates from a \ac{mmw} localization algorithm, in order to relieve the training dataset collection effort. Our performance evaluation suggests that the \ac{nn} model achieves a high degree of accuracy when trained with perfect data, in spite of its shallowness, even in the presence of significant \ac{aoa} estimation errors. Such accuracy does not significantly degrade when trained with imperfect location labels. 
Results show sub-meter localization accuracy in $\approx90\%$ of the scenarios with large \ac{aoa} errors. 
Future work includes the experimental validation of the proposed method using \ac{mmw} \ac{cots} devices.

\section*{Acknowledgment}

This project has received funding from the EU's Framework Programme for Research and Innovation Horizon~2020 under Grant Agreement No. 861222 (EU H2020 MSCA MINTS).

\bibliographystyle{IEEEtran}
\bibliography{IEEEabrv,references}

\end{document}